\documentclass[twocolumn]{aastex701}
\usepackage{amsmath}
\usepackage{hyperref}
\usepackage{graphicx}
\usepackage{booktabs}
\usepackage{diagbox}
\usepackage{xcolor}
\usepackage{color}
\usepackage{bold-extra}
\usepackage{enumitem}
\usepackage{url}
\usepackage{tabularx}

\newcommand{\I}{\text{\uppercase\expandafter{\romannumeral 1}}}
\newcommand{\II}{\text{\uppercase\expandafter{\romannumeral 2}}}
\newcommand{\III}{\text{\uppercase\expandafter{\romannumeral 3}}}
\newcommand{\IV}{\text{\uppercase\expandafter{\romannumeral 4}}}

\newcommand{\ha}{\rm{H$\alpha$}}
\newcommand{\hb}{\rm{H$\beta$}}

\shorttitle{H$\alpha$ LF from CANUCS}
\shortauthors{Zhou et al.}

\begin{document}

\title{A Comprehensive Analysis of the H$\alpha$ Luminosity Functions at $z\sim 1.3$ and $z\sim 2.0$ in the CANUCS Lensing Fields with JWST NIRISS}

\author[orcid=0009-0006-1255-9567]{Qianqiao Zhou}
\affiliation{School of Astronomy and Space Science, University of Chinese Academy of Sciences (UCAS), Beijing 100049, China}
\email[]{zhouqianqiao24@mails.ucas.ac.cn}  

\author[0009-0005-3823-9302]{Yuxuan Pang}
\affiliation{School of Astronomy and Space Science, University of Chinese Academy of Sciences (UCAS), Beijing 100049, China}
\email[]{pangyuxuan@ucas.ac.cn}

\author[orcid=0000-0002-9373-3865]{Xin Wang}
\affiliation{School of Astronomy and Space Science, University of Chinese Academy of Sciences (UCAS), Beijing 100049, China}
\affiliation{National Astronomical Observatories, Chinese Academy of Sciences, Beijing 100101, China}
\affiliation{Institute for Frontiers in Astronomy and Astrophysics, Beijing Normal University, Beijing 102206, China}
\email[]{xwang@ucas.ac.cn}

\author[0000-0002-3254-9044]{Karl Glazebrook}
\affiliation{Centre for Astrophysics and Supercomputing, Swinburne University of Technology, Hawthorn, VIC 3122, Australia}
\email[]{}

\author[0009-0007-6655-366X]{Shengzhe Wang}
\affiliation{School of Astronomy and Space Science, University of Chinese Academy of Sciences (UCAS), Beijing 100049, China}
\affil{National Astronomical Observatories, Chinese Academy of Sciences, Beijing 100101, China}
\email[]{}

\author[orcid=0009-0004-7133-9375]{Hang Zhou}
\affiliation{School of Astronomy and Space Science, University of Chinese Academy of Sciences (UCAS), Beijing 100049, China}
\email[]{zhouhang23@mails.ucas.ac.cn}

\author[orcid=0009-0008-9411-2770]{Pengfei Ren}
\affiliation{School of Astronomy and Space Science, University of Chinese Academy of Sciences (UCAS), Beijing 100049, China}
\email[]{renpengfei20@mails.ucas.ac.cn}

\author[0009-0003-1463-4397]{Shamuhawu Hailanhazi}
\affil{School of Astronomy and Space Science, University of Chinese Academy of Sciences (UCAS), Beijing 100049, China}
\email[]{}


\author[0000-0003-3108-0624]{Peter J. Watson}
\affiliation{Universit\'e C\^ote d'Azur, Observatoire de la Côte d’Azur, CNRS, Laboratoire Lagrange, 06000 Nice, France}
\email[]{}


\author[0000-0003-4570-3159]{Nicha Leethochawalit}
\affiliation{National Astronomical Research Institute of Thailand (NARIT), Mae Rim, Chiang Mai, 50180, Thailand}
\email[]{}

\author[0000-0001-6919-1237]{Matthew A. Malkan}
\affiliation{Department of Physics and Astronomy, University of California, Los Angeles, 430 Portola Plaza, Los Angeles, CA 90095, USA}
\email[]{}

\author[0000-0003-2804-0648]{Themiya Nanayakkara}
\affiliation{Sydney Institute for Astronomy, School of Physics, The University of Sydney, Sydney, NSW 2006, Australia}
\email[]{}

\author[0000-0002-7350-6913]{Xue-Bing Wu}
\affiliation{Department of Astronomy, School of Physics, Peking University, Beijing 100871, People's Republic of China}
\affiliation{Kavli Institute for Astronomy and Astrophysics, Peking University, Beijing 100871, People's Republic of China}
\email[]{}

\author[0000-0003-1718-6481]{Hu Zhan}
\affil{National Astronomical Observatories, Chinese Academy of Sciences, Beijing 100101, China}
\affil{Kavli Institute for Astronomy and Astrophysics, Peking University, Beijing 100871, China}
\email[]{}

\correspondingauthor{Yuxuan Pang, Xin Wang}
\email{pangyuxuan@ucas.ac.cn, xwang@ucas.ac.cn}
\begin{abstract}

We present H$\alpha$ luminosity function (LF) measurements at redshifts $z\sim1.3$ and $z\sim2.0$. We adopt JWST grism data in five lensing cluster fields and their parallel fields from the CANUCS program: ABELL-370, MACS-J0416, MACS-J0417, MACS-J1149, MACS-J1423. Based on emission lines spectroscopically identified in the F115W, F150W and F200W filters, our final sample consists of 762 H$\alpha$ emitters, including 433 at $1.1<z<1.5$ and 329 at $1.8<z<2.2$. Through detailed effective volume and completeness analysis for each source, we construct the H$\alpha$ LF in two redshift bins. Thanks to the sensitivity of NIRISS WFSS and gravitational lensing magnification, our sample reaches intrinsic H$\alpha$ luminosities down to $L_{\rm H\alpha}\sim10^{40.6}~\rm erg~s^{-1}$ at $z\sim1.3$ and $L_{\rm H\alpha}\sim10^{41}~\rm erg~s^{-1}$ at $z\sim2.0$ with completeness larger than 80\%. The larger survey volume and coverage of multiple independent sightlines help mitigate cosmic variance. 
After accounting for cosmic variance, we constrain the faint-end slopes to be $\alpha_{\mathrm{H}\alpha}=-1.51^{+0.06}_{-0.04}$ at $z\sim1.3$ and $-1.42^{+0.09}_{-0.06}$ at $z\sim2.0$, indicating a mild evolution between the two redshift bins. We find a negligible contribution from bright active galactic nuclei in our sample. We estimate integrated cosmic star formation rate densities of $0.103^{+0.008}_{-0.007}~M_{\odot}~\rm yr^{-1}~Mpc^{-3}$ at $z\sim1.3$ and $0.149^{+0.016}_{-0.013}~M_{\odot}~\rm yr^{-1}~Mpc^{-3}$ at $z\sim2.0$. 
The emission-line samples presented here will enable further detailed studies of galaxy properties including metallicities. The methodology presented here can be readily applicable to other JWST slitless spectroscopic datasets and future wide-field slitless surveys, including those from Euclid, Roman, and the Chinese Space Station Telescope.
\end{abstract}



\section{Introduction} \label{sec: intro}
Around 50\% of the stellar mass currently observed in galaxies was formed over approximately 3.5 billion years, from redshifts $z\sim3$ to $z\sim1$, a period referred to as cosmic noon. \citep{SFG_review, z0_to_z4, 2010ApJ...709.1018V, 2011A&A...528A..35M, 2013A&A...556A..55I, 2013ApJ...777...18M, 2014ARA&A..52..415M}. Previous studies have shown that the strong relationship between the star formation rate (SFR) and stellar mass of galaxies (the Main Sequence), observed in the local universe, was already established by at least $z \sim 2$ \citep{2007ApJ...670..156D, 2007A&A...468...33E, 2007ApJ...660L..43N}. Additionally, other scaling relations including those related to size, kinematics, and the content of metals and gas, are also followed by star-forming galaxies up to this redshift \citep{2014ApJ...788...28V, 2017ApJ...842..121U, 2019A&ARv..27....3M, 2020ARA&A..58..157T}.

Measuring the cosmic star-formation rate during cosmic noon is a crucial step in reconstructing the star-formation history (SFH). Several SFR tracers have been developed and utilized in previous studies \citep{2012ARA&A..50..531K, 2014ARA&A..52..415M}. Compared to UV radiation and the Ly$\alpha$ line, the \ha\ emission line, which traces ionizing photons from young massive stars \citep{2022MNRAS.513.2904T}, is less affected by dust attenuation and experiences relatively simpler radiation transport.

Therefore, many projects aim to use the \ha\ luminosity function to reveal the star formation rate density during cosmic noon \citep{2014ARA&A..52..415M}. These projects typically use three approaches: narrowband photometry with ground-based telescopes to select \ha\ emitters \citep{1998ApJ...506..519T, 2012PASP..124..782L, 2013MNRAS.428.1128S}, spectroscopic follow-up after photometric selection in certain band \citep{1999MNRAS.306..843G, 2007IAUS..235..417R, 2013MNRAS.433.2764G}, or Hubble Space Telescope (HST) spectroscopic surveys to identify \ha\ over a wider redshift range \citep{2009ApJ...696..785S, 2023ApJ...943....5N}. The slitless spectroscopic technique employed by space telescopes avoids limitations such as atmospheric interference and slit losses, and is particularly efficient at detecting high-equivalent-width emission-line galaxies. It enables the acquisition of galaxy spectra across the entire field of view that satisfy flux and redshift constraints, without requiring pre-selected targets \citep{2023ApJ...943....5N, 2025ApJ...987..186F, 2026A&A...708A..33M,2019ApJ...875..152B}. Previous projects such as GRAPES \citep{2004ApJS..154..501P}, WISP \citep{2010ApJ...723..104A}, and 3D-HST \citep{2012ApJS..200...13B} have used HST grisms to establish the first space telescope-based emission-line galaxy (ELG) samples.

Compared to HST/WFC3 and ACS, JWST WFSS significantly improves both the observation depth and wavelength coverage, allowing for better constraints on the faint end of the \ha\ luminosity function and extending the redshift range to $z > 2$ \citep{2025A&A...699A.225W}. 
Recent surveys such as FRESCO \citep{2023MNRAS.525.2864O}, CONGRESS (Sun et al., in prep.), and MAGNIF \citep{2025ApJ...987..186F} have advanced statistical studies of \ha\ emitters at high redshift, providing constraints on the cosmic star formation rate density (CSFRD) and insights into galaxy evolution and large-scale structure \citep{2025A&A...694A.178C, 2023ApJ...953...53S, 2026ApJ...997..207L}. These studies primarily use NIRCam grism observations, which probe \ha\ at $z \approx 3.7$--$6.6$. At lower redshifts, previous \ha\ LF measurements at $z \sim 1.3$ mainly constrained the bright end \citep{2013MNRAS.428.1128S, 2023ApJ...943....5N}, while those at $z \sim 2$ relied on narrowband surveys before JWST \citep{2010A&A...509L...5H}. Using JWST/NIRISS slitless spectroscopy from the GLASS-JWST Early Release Science Program, \citet{Pang_} (here after P26) measured the \ha\ LFs at $z \sim 1.3$ and $z \sim 2.0$, reaching $\sim10$ times deeper than previous grism surveys. Their measurements substantially improve constraints on the faint end, while yielding LFs and CSFRDs broadly consistent with previous studies.

We note that estimates of galaxy population properties are inherently affected by cosmic variance, i.e., the uncertainty in measured galaxy number densities caused by large-scale density fluctuations in the Universe \citep{2011ApJ...731..113M}. This effect becomes especially significant at high redshift, where galaxies are expected to exhibit stronger clustering than the underlying dark matter distribution \citep{Kauffmann_1999, 1999MNRAS.305L..21B, 2004ApJ...609..525C, 2010ApJ...710..903M}. Moreover, rare starburst galaxies and large-scale overdense regions may also contribute significantly to the measured star formation rate density \citep{rodighiero2011lesser,williams2024galaxies,sun2025spectroscopic,martis2026new}. In this context, estimating the LF across multiple fields and a large survey volume helps mitigate uncertainties arising from cosmic variance.

In this work, we measure the H$\alpha$ luminosity function (LF) and  CSFRD at $z \sim 1.3$ and $z \sim 2.0$ using JWST/NIRISS wide-field slitless spectroscopy from the CANUCS survey. We adopt the framework of P26 to calculate survey volume and completeness. Leveraging JWST’s sensitivity and gravitational lensing from five cluster fields, we detect H$\alpha$ emitters down to $L_{H\alpha} \sim 10^{40.6},\mathrm{erg,s^{-1}}$ at $z \sim 1.3$ and $\sim 10^{41.0},\mathrm{erg,s^{-1}}$ at $z \sim 2.0$, corresponding to star formation rates of $\sim 0.4$ and $\sim 1.0\ M_\odot\ \mathrm{yr^{-1}}$, respectively.

This paper is organized as follows. Section \ref{sec: data} describes the observations and data reduction, and Section \ref{sec: sample} outlines the selection of the H$\alpha$ emitter sample. Section \ref{sec: method} describes the adopted lens models, the identification of multiply imaged systems, and the methods for estimating survey volumes, sample completeness, flux bias, and dust attenuation. Section \ref{sec: LF} presents the H$\alpha$ luminosity functions and their Schechter fits. Section \ref{sec: discussion} discusses cosmic variance, compares the faint-end slopes with previous studies, and derives the CSFRD under different dust attenuation assumptions. Finally, Section \ref{sec: summary} summarizes our conclusions. Throughout this work, we assume a $\Lambda$CDM cosmology with $\Omega_{m} = 0.3$, $\Omega_{\Lambda} = 0.7$, and $H_0 = 70\ \mathrm{km\ s^{-1}\ Mpc^{-1}}$. All cosmological distances are computed using the Cosmology Calculator \citep{cosmic_cal}.

\section{Observations and Data Reduction} \label{sec: data}
We make use of data from the Canadian NIRISS Unbiased Cluster/Technicolor Survey \citep[CANUCS/Technicolor; GTO-1208, GO-3362;][]{2022PASP..134b5002W, 2026ApJS..282....3S}. This program observes five strong-lensing galaxy clusters along with their parallel fields (A370, MACS~J0416.1-2403, MACS~J1149.6+2223, MACS~J1423.8+2404, and MACS~J0417.5-1154), covering a total area of $\sim 100\ \mathrm{arcmin^2}$. The dataset includes NIRCam imaging, NIRISS slitless spectroscopy, and NIRSpec prism multi-object spectroscopy. A full description of the observing strategy is provided in \citet{2022PASP..134b5002W}.
In summary, each cluster is observed with NIRISS using two orthogonal grisms (GR150C and GR150R) and three filters (F115W, F150W, and F200W), with a total exposure time of 9.7 ks per configuration. These low-resolution grisms ($R \sim 150$), spanning $1.0$–$2.3\ \mu$m, allow for the simultaneous detection of H$\alpha$ and [O III] $\lambda5007$ over a redshift range of approximately $z \sim 1.1$ to $2.5$. There are no prominent peaks in the two redshift bins, although a mild concentration is seen at $1.1<z<1.5$, with a peak at $z = 1.197$.

Our data reduction follows the procedure outlined in P26 and is broadly consistent with \citet{2022ApJ...938L..13R}. We adopt the \texttt{jwst\_1413.pmap} reference files, which incorporate updated calibrations for NIRISS WFSS background treatment, and use the grism analysis software \texttt{GRIZLI} \citep[][version 1.12]{grizli}. For both GR150C and GR150R, we adopt the NIRISS wavelength calibration configurations of \citet{matharu_brammer2022} for the F115W, F150W, and F200W filters. We begin with count-rate products from the \texttt{Detector1} stage of the STScI pipeline, where detector-level corrections (e.g., bad pixel masking, cosmic ray rejection, dark subtraction, and ramp fitting) have already been applied. These data are then processed with the \texttt{GRIZLI} preprocessing pipeline, which carries out astrometric alignment, WCS registration, flat-field correction, sky subtraction, and pixel resampling to generate fully reduced exposures and mosaics for both direct imaging and WFSS observations. 
All images are registered to the Legacy Surveys DR10 \citep{2019AJ....157..168D} and Gaia DR3 \citep{2023A&A...674A...1G} reference frames. For the final mosaics, we adopt a pixel scale of 30 mas to enable spatially resolved analyses of emission-line structures \citep{2022ApJ...938L..13R}. Source detection and photometry are performed with the \texttt{SEP} package on the combined F115W, F150W, and F200W mosaic, from which segmentation maps are generated and used as inputs for forward modeling.

For sources with $m_{\mathrm{AB}} < 29$, we model the contamination by fitting a polynomial function to their spectra. Following \citet{2022ApJ...938L..13R}, we use a combination of the contamination level and flux uncertainty as pixel-wise weights for spectral extraction and modeling. We adopt flexible stellar population synthesis templates that include emission-line complexes with
fixed ratios \citep{2009ApJ...699..486C, 2010ApJ...712..833C, 2023ApJS..266...13S} to perform forward modeling of the 2D grism spectra. The corresponding 1D spectra are optimally extracted from the 2D data using the morphological model of each source to define the position and spatial profile of its spectral trace.

\begin{figure*}[!t]
    \centering
    \includegraphics[width=\textwidth]{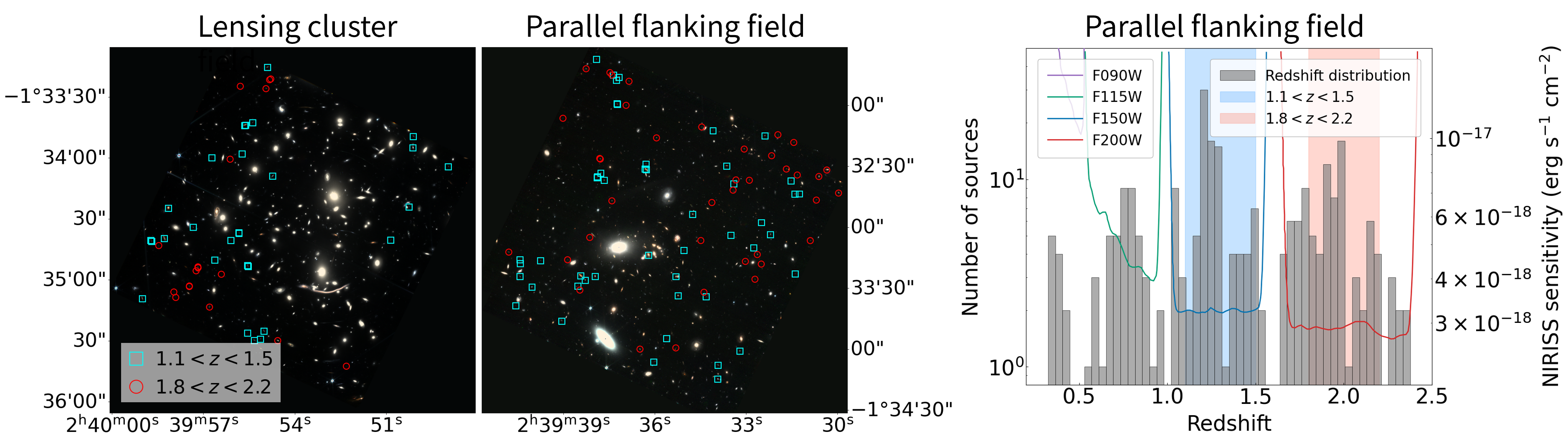}
    \vspace*{-1em}
    \caption{
    \textbf{Left:} Spatial distribution of \ha\ emitters at $1.1<z<1.5$ (cyan squares) and $1.8<z<2.2$ (red circles) in the lensing cluster field of ABELL~370. The background shows a color-composite image, with F115W, F150W, and F200W mapped to blue, green, and red, respectively. 
    \textbf{Middle:} Same as the left panel, but for the parallel field of ABELL~370. 
    \textbf{Right:} Redshift distribution of the selected \ha\ emitters in ABELL~370, with the selection criteria described in Section~\ref{sec: sample}. The blue and red shaded regions indicate the two redshift bins used in our analysis ($1.1<z<1.5$ and $1.8<z<2.2$, respectively). The colored curves show the JWST/NIRISS flux sensitivities in the F090W, F115W, F150W, and F200W bands, evaluated at the observed wavelength of \ha\ as a function of redshift.
    }
    \label{fig:distribution}
\end{figure*}

\section{\ha\ Emitter Samples} \label{sec: sample}
We adopt the sample selection criteria from P26. Due to the limited spectral resolution of NIRISS, the observed H$\alpha$ flux is blended with the [N II] doublet. We use the terms ``H$\alpha$ sample", ``H$\alpha$ luminosity ($L_{H\alpha}$)", and ``H$\alpha$ LF" to refer to these combined H$\alpha$+[N II] measurements following the convention of \citet{2023ApJ...943....5N, 2013ApJ...779...34C}.

The automated selection aims to assess the fitted redshift and quality derived from the fitting of 2D grism spectra and broadband photometry. The selected galaxies satisfy the following criteria:
\begin{enumerate}
\setlength{\itemsep}{0pt}
\setlength{\parskip}{0pt}
\setlength{\topsep}{2pt}    
\setlength{\partopsep}{2pt} 
\item A total infrared magnitude of \\
\mbox{$M_{\rm F115W+F150W+F200W} < 27$}, ensuring sufficient signal-to-noise ratio (SNR) in the grism spectra.
\item A 68\% confidence interval of $\Delta z = |z_{\rm grism,u68} - z_{\rm grism,l68}| < 0.005(1+z)$, indicating well-constrained redshift estimates.
\item A redshift range of $0.54 < z_{\rm grism} < 2.39$, such that H$\alpha$ lies within the wavelength coverage of the F115W, F150W, or F200W filters.
\item An acceptable fit quality with $\chi^2 < 3$.
\item 
H$\alpha$ flux with $\mathrm{SNR}_{\mathrm{H}\alpha} > 5$, measured from the 1D spectra.
\end{enumerate}

We find that the \ha\ signal-to-noise ratio (SNR) derived from the \texttt{GRIZLI} SED fitting is generally larger than that obtained from direct integration of the 1D spectrum within a window of $\pm \rm FWHM$ around the line center. We therefore require that both SNR measurements satisfy the fifth criterion described above. 

Following \citet{2014ApJ...790..113Z} and \citet{2019ApJ...875..152B}, we compiled relevant information for each source and conducted a visual inspection based on the reliability of the redshift measurements and the presence of H$\alpha$ and other ancillary emission lines (\hb, [O \III] $\lambda5008$). This inspection incorporated multi-band imaging, 1D and 2D spectra with best-fit models, redshift probability distributions, and uncontaminated line maps. We also removed one additional source in MACS-J0417 due to the presence of a broad H$\alpha$ emission component. Sources exhibiting severe contamination or lacking clear emission-line detections in the grism data were subsequently excluded. Approximately 40\% of the inspected sources were retained as reliable H$\alpha$ emitters based on visual inspection.

To obtain the most reliable \ha\ detections, we restrict our analysis to two redshift intervals, $1.1<z<1.5$ and $1.8<z<2.2$, where the \ha\ emission line falls near the center of the F150W and F200W filters, respectively. 
We exclude sources with \ha\ falling in the F115W band, where \hb\ and [O \III] $\lambda5008$ lie outside our wavelength coverage. Alternative diagnostic lines, such as Paschen $\beta$ and [S \III], are often weaker and detected at lower SNR, making secure redshift confirmation more difficult. Real \ha\ emitters may therefore be rejected by our redshift-confirmation criteria, reducing the completeness of the sample at lower redshifts.
In addition, lower-redshift sources tend to exhibit more extended morphologies, which introduce greater uncertainties in line detection and reduce the overall sample completeness. Figure~\ref{fig:distribution} shows the spatial distribution of the final \ha\ emitter sample in ABELL~370, along with the redshift distribution of sources prior to applying the redshift cuts.

\begin{figure*}[!t]
    \centering
\includegraphics[width=\textwidth]{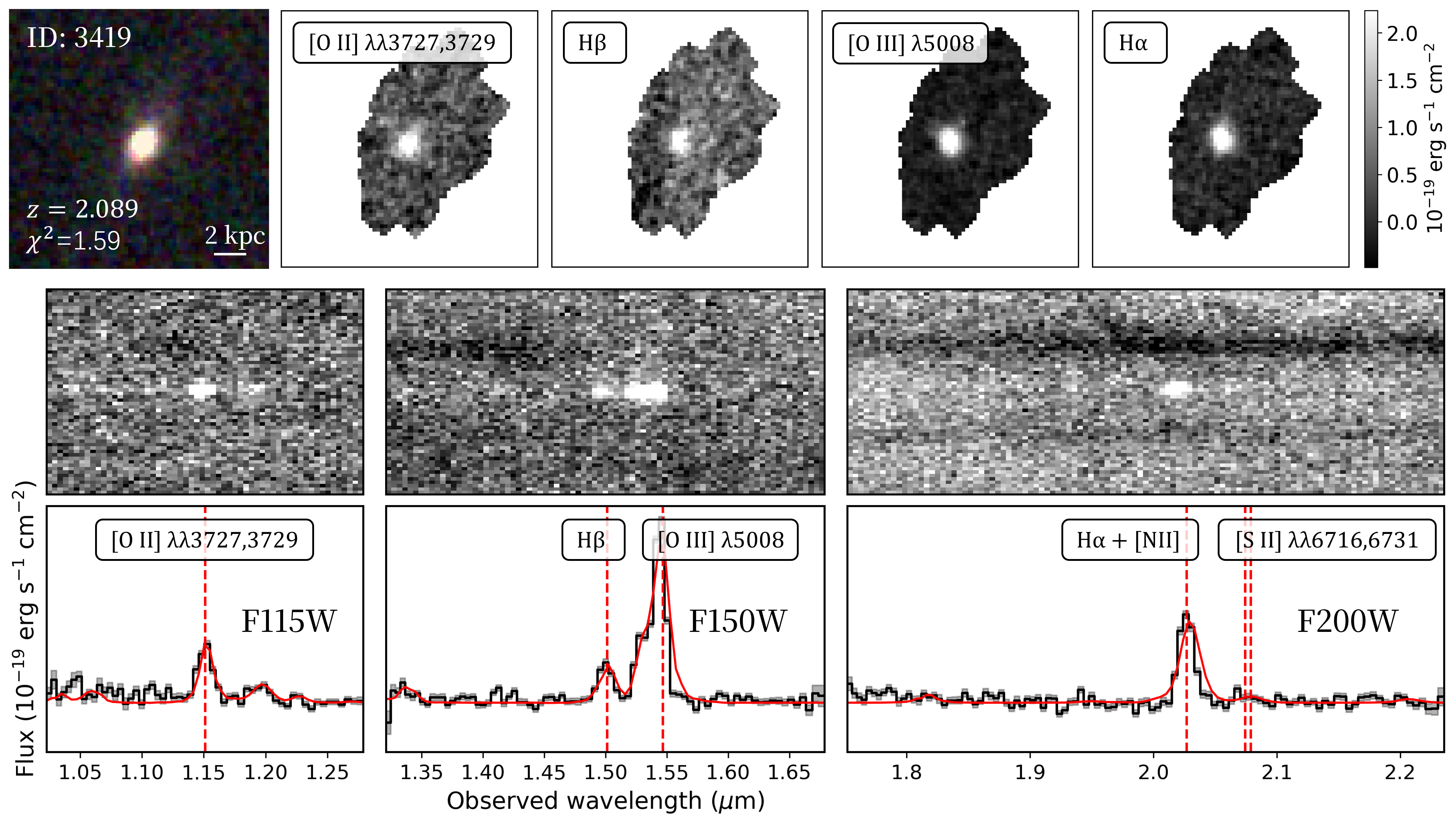}
    \caption{2D emission-line maps, 2D spectra, and 1D spectra for a representative source in MACS~J0416. \textbf{Top:} Color-composite cutout (R: F200W, G: F150W, B: F115W) and surface-brightness maps of the emission lines ([O \II], H$\beta$, [O \III], and H$\alpha$). The redshift, best-fit $\chi^2$, and physical scale are indicated in the left panel. \textbf{Middle:} continuum-subtracted 2D grism spectra obtained in the F115W, F150W, and F200W filters. \textbf{Bottom:}  continuum-subtracted 1D spectrum, with the $1\sigma$ uncertainty shown as gray shading. The red curves represent the forward-modeling results, and the expected wavelengths of prominent emission lines are marked by red dashed vertical lines.}
    \label{fig:stamp}
\end{figure*}

Figure~\ref{fig:stamp} shows an example from our sample. The top panel presents the color-composite cutout along with the segmentation maps of [O \II] $\lambda\lambda 3727,3729$, \hb, [O \III] $\lambda 5008$, and \ha. The middle and bottom panels display the corresponding 2D and 1D spectra. This source, at redshift $z \approx 2.089$, exhibits \ha\ in F200W, \hb\ and [O \III] $\lambda 5008$ in F150W, and [O \II] $\lambda\lambda 3727,3729$ in F115W, which together help to further confirm the redshift.

\section{Methodology} \label{sec: method}

\subsection{Lens Models and Multiply Imaged Systems} \label{subsec: lensmodel}
Our sample consists of five massive galaxy cluster lensing fields and their associated parallel fields. To correct for gravitational lensing magnification and identify potential multiple images, we adopt the publicly available CANUCS DR1 lens models \citep{2026ApJS..282....3S}, which are constrained using the parametric lens modeling software \texttt{Lenstool} \citep{1996ApJ...471..643K, 2007NJPh....9..447J, 2009MNRAS.395.1319J}. Magnification factors are derived from the corresponding deflection maps. We note that CANUCS DR1 also provides catalogs of lensing magnification factors. We find no significant differences between the results derived from the deflection maps and those reported in these catalogs.

To compute the survey volume by integrating over the depth map for each source, we generate magnification maps across a grid of redshifts, with a step size of $\Delta z = 0.01$ over the ranges $1.1 < z < 1.5$ and $1.8 < z < 2.2$.

Using the deflection maps derived from the lensing models described above, together with the redshifts provided by \texttt{GRIZLI}, we mapped all sources from the image plane to the source plane and derived their corresponding source-plane positions. We then searched for candidate multiply imaged systems by identifying sources with consistent redshifts that map to nearly the same positions in the source plane. This analysis identified one candidate doubly imaged system in ABELL 370, four in MACS J0416, two in MACS J0417, and two in MACS J1149. 
To validate our identification procedure, we cross-matched our candidates against the CANUCS DR1 catalogs of secure multiply imaged systems with spectroscopic redshifts. These reference catalogs are available only for ABELL 370 and MACS J0416. The comparison confirms that our source-plane matching approach effectively recovers known multiply imaged systems.

\subsection{Survey Volume} \label{subsec: Vmax}
Following the method of P26, we estimate the survey volume and completeness for each source. 
We divide the CANUCS field of view into $2'' \times 2''$ grids and extract 1D grism spectra at each grid position in each field and redshift bin. The extraction procedure is identical to that implemented in \texttt{GRIZLI} for real emitters. 
To account for contamination, we incorporate the contamination model into the uncertainties of the 1D spectra, weighted according to \citet{2022ApJ...938L..13R}. 
This approach effectively propagates flux contamination from nearby bright sources into increased spectral uncertainties at the corresponding wavelengths.

We adopt the measured FWHM of the observed \ha\ emitters as the integration window for the \ha\ flux and for estimating the corresponding errors at each spatial and redshift grid point. We divide the two redshift bins ($1.1<z<1.5$ and $1.8<z<2.2$) into steps of $\Delta z = 0.01$, and compute the 1D spectral errors within a window of $\pm \mathrm{FWHM}$ centered on the line position at each simulated redshift. 

For each source with a measured \ha\ luminosity, we compare its magnification-corrected intrinsic flux to the error map (also corrected for the local magnification) at each redshift step. This allows us to determine the region in the image plane where the lensed flux would exceed the local $5\sigma$ detection threshold at each redshift step. At the same time, we derive the corresponding survey area in the source plane by correcting for the magnification at each position. Then we integrated the source-plane area across each redshift bin and obtained the total survey volume. Table~\ref{tab:LF} lists the mean relative survey volume, $\langle V_{\max}/V_{\max}(L_{\max})\rangle$, in each luminosity and redshift bin. For each galaxy, we divide its survey volume by $V_{\max}(L_{\max})$, the largest $V_{\max}$ among all galaxies in the same pointing and redshift bin and then calculate the mean ratio in each luminosity and redshift bin, with equal weight given to each contributing pointing.

\begin{figure*}[!t]
    \centering
    \includegraphics[width=\textwidth]{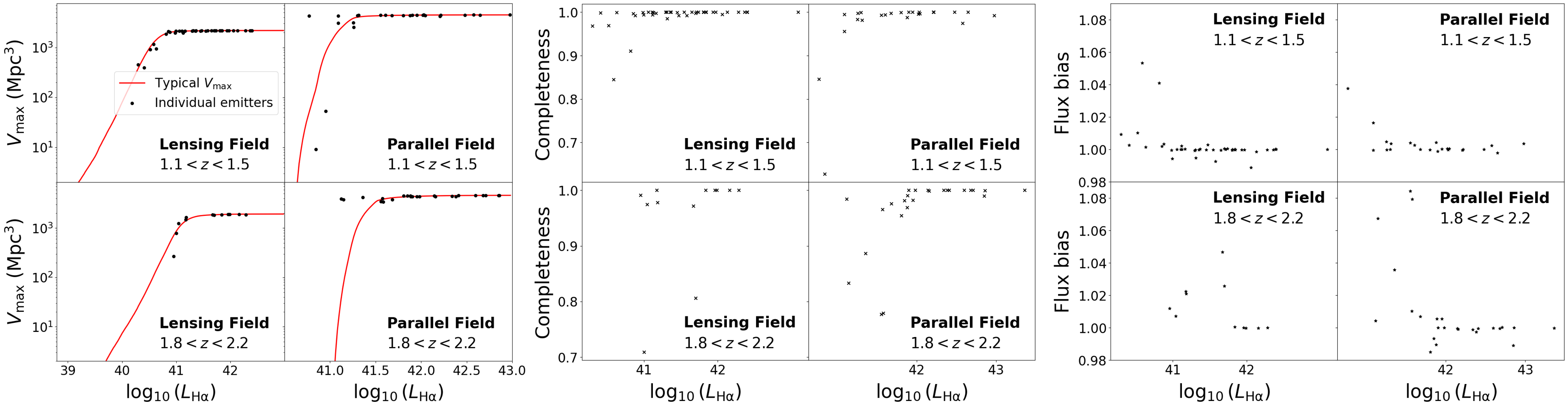}
    \caption{$\rm V_{max}$, completeness, and flux bias versus intrinsic \ha\ luminosity (corrected by magnification) for the lensing cluster field (first, third, and fifth columns) and the parallel field (second, fourth, and sixth columns) of MACS J1149. The top and bottom rows correspond to the redshift bins $1.1<z<1.5$ and $1.8<z<2.2$, respectively. The red curves in the $\rm V_{max}$–\ha\ luminosity panels are computed by integrating the survey area that satisfies the SNR selection at each luminosity grid, as defined by the depth maps, together with the corresponding luminosity distance over the relevant redshift range.
}
    \label{fig:placeholder}
\end{figure*}

Notably, we treat the lensing cluster fields and the parallel fields as distinct survey regions, so the full sample is drawn from ten independent areas. Each source is assigned a survey volume limited to its own region (either the lensing cluster field or the parallel field). Figure~\ref{fig:placeholder} presents the $V_{\max}$ results for MACS--J1149 as an illustrative example. The red curve is computed using the same procedure as for the real sources, but adopting the median FWHM as the integration window. A significant discrepancy emerges between these two $V_{\max}$ estimates, likely driven by the variation in the morphology and extent of galaxies at cosmic noon, as reflected in their \ha\ line widths.

\subsection{Sample Completeness \& Flux Bias} \label{subsec: completeness}
A robust determination of the LF requires accounting for the effect of the flux fluctuation induced by observational noise.
We therefore incorporated these effects through source-specific completeness and flux-bias simulations. Following the Monte Carlo framework of P26, we performed injection simulations for individual H$\alpha$ emitters, explicitly accounting for spatial variations in the local noise properties across each field. The main steps of the procedure are summarized as follows:

\begin{enumerate}
\setlength{\itemsep}{0pt}
\setlength{\parskip}{0pt}
\setlength{\topsep}{2pt}    
\setlength{\partopsep}{2pt} 

\item For each source, we randomly selected 200 positions within the area of the corresponding depth map where that source would satisfy the detection criterion, and extracted the associated 1D spectra using the true segmentation map of the source.

\item For each source, we generated simulated total fluxes by drawing from a Gaussian distribution centered on the observed flux, with the width set by its measurement uncertainty. Both the simulated fluxes and their uncertainties were adjusted according to the magnification at each mock position. The one-dimensional H$\alpha$ spectral profile at each mock position was then scaled by the ratio of simulated total flux to the magnification-corrected flux before Gaussian sampling.

\item At each mock position, we generated 100 mock spectra for each source by drawing the flux in each wavelength from a Gaussian distribution centered on the rescaled one-dimensional spectrum, with the width set by the corresponding one-dimensional error spectrum at that position.

\item For each source, the completeness was defined as the fraction of simulated realizations satisfying the $5\sigma$ detection criterion among the 20,000 mock spectra generated from 200 positions and 100 realizations per position.

\end{enumerate}
The mean completeness values averaged across all fields in each luminosity bin are listed in Table~\ref{tab:LF}.

\begin{deluxetable}{ccccc}
\tablecaption{\centering H$\alpha$ Luminosity Functions
\label{tab:LF}}
\tabletypesize{\footnotesize}
\tablehead{
\colhead{$\log L_{\rm H\alpha}$} & \colhead{$N$} & \colhead{$\langle c\rangle$} &
\colhead{$\langle V_{\max}/V_{\rm max}(L_{max})\rangle$} &
\colhead{$\log \langle\Phi_{\rm H\alpha}\rangle$} \\
\colhead{($\rm erg\,s^{-1}$)} & & & &
\colhead{($\rm Mpc^{-3}\,dex^{-1}$)} \\
\colhead{(1)} & \colhead{(2)} & \colhead{(3)} & \colhead{(4)} & \colhead{(5)}
}
\startdata
\multicolumn{5}{c}{$z\sim1.3$} \\
\hline
$40.8\pm0.2$ & 22  & 0.97 & 0.74 & $-1.99^{+0.09}_{-0.09}$ \\
$41.2\pm0.2$ & 132 & 0.99 & 0.87 & $-1.99^{+0.05}_{-0.05}$ \\
$41.6\pm0.2$ & 98  & 1.00 & 0.95 & $-2.26^{+0.06}_{-0.06}$ \\
$42.0\pm0.2$ & 68  & 1.00 & 0.99 & $-2.59^{+0.07}_{-0.07}$ \\
$42.4\pm0.2$ & 38  & 1.00 & 1.00 & $-2.86^{+0.08}_{-0.09}$ \\
$42.8\pm0.2$ & 7   & 1.00 & 1.00 & $-3.84^{+0.21}_{-0.23}$ \\
\hline
\multicolumn{5}{c}{$z\sim2.0$} \\
\hline
$41.2\pm0.2$ & 24  & 0.93 & 0.80 & $-2.20^{+0.10}_{-0.10}$ \\
$41.6\pm0.2$ & 82  & 0.97 & 0.86 & $-2.41^{+0.07}_{-0.07}$ \\
$42.0\pm0.2$ & 104 & 1.00 & 0.98 & $-2.41^{+0.06}_{-0.06}$ \\
$42.4\pm0.2$ & 44  & 1.00 & 0.99 & $-2.80^{+0.09}_{-0.09}$ \\
$42.8\pm0.2$ & 21  & 1.00 & 0.99 & $-3.21^{+0.12}_{-0.12}$ \\
\enddata
\tablecomments{
Column (1): logarithm of the H$\alpha$ luminosity at the center of each luminosity bin.
Column (2): number of H$\alpha$ emitters in each bin.
Column (3): mean completeness of H$\alpha$ emitters in each bin.
Column (4): mean ratio between the survey volume of the H$\alpha$ emitters and the max survey volume in all fields.
Column (5): logarithmic H$\alpha$ number density, with the uncertainty including both Poisson uncertainty and cosmic variance.
}
\end{deluxetable}

We additionally quantified Eddington biases introduced by the detection and extraction procedure. For each simulated source, the recovered fluxes of successfully detected mock emission lines were compared with their corresponding input fluxes. The median ratio between the mocked and input fluxes was adopted as the flux correction factor, hereafter referred to as the flux bias, and was used to correct the measured H$\alpha$ fluxes.

\subsection{Dust Correction} \label{dust}
To infer star formation rates from the observed $H\alpha +[\mathrm{N\,II}]$ flux, we need to account for dust attenuation of H$\alpha$ luminosity. We describe the correction for the $[\mathrm{N\,II}]$ contribution in Section~\ref{sec: LF}. For our main LF and CSFRD analyses in Sections~\ref{sec: LF} and~\ref{subsec: CSFRD}, we first adopt a fiducial attenuation of $A_{\mathrm{H}\alpha}=1\,\mathrm{mag}$, consistent with values adopted in some previous studies \citep{Pang_,2013MNRAS.428.1128S,sobral2012star,2010A&A...509L...5H,khostovan2020large}.

We then perform the following tests to better constrain the dust attenuation of our sample. We utilize stacked maps of the \ha\ and \hb\ emission lines, which significantly improve the signal-to-noise ratio (SNR). The stacking procedure follows \citet{2026ApJ..1005L..20R}, and is briefly summarized here. 
We visually inspected the H$\beta$ maps and excluded sources with clear contamination from [O \III] $\lambda5007$ emission. For each field and redshift bin, we stacked the H$\alpha$ and H$\beta$ maps separately by taking the median at each pixel. We integrated both stacks within segmentation masks and computed the Balmer decrement as the ratio of the integrated fluxes. We estimated its uncertainty from 1,000 Monte Carlo realizations, perturbing each input pixel by its measurement uncertainty and repeating the stacking and integration.
We then used the resulting H$\alpha$/H$\beta$ ratio to derive the Balmer decrement and constrain the H$\alpha$ attenuation, $A_{\mathrm{H}\alpha}$. 

The resulting H$\alpha$/H$\beta$ ratios for each field and redshift bin are listed in Table~\ref{tab: LF_para_individual}. Assuming Case~B recombination with an electron temperature of $T_e = 10^4 \rm K$ and an electron density of $n_e = 100 \rm cm^{-3}$, the intrinsic \ha/\hb\ ratio is taken to be 2.86 \citep{1989agna.book.....O}. We can estimate the $A_{H\beta}$ with following equations: 
\begin{equation}
\begin{split}
\log[L_{\rm H\alpha}/({\rm erg~s^{-1}})] =  \log(2.86) + \log[L_{\rm H\beta}/({\rm erg~s^{-1}})]
\end{split}
\end{equation}
\begin{equation}
\begin{split}
\log[L_{\rm H\beta}/({\rm erg~s^{-1}})] = \log[L_{\rm H\beta,~obs}/({\rm erg~s^{-1}})] + 0.4A_{\rm H\beta}
\end{split}
\end{equation}
Following the extinction curve of \citet{calzetti2000}, we convert $A_{H\beta}$ to $A_{H\alpha}$. 

We also stacked the line maps across lensing fields and parallel fields separately to derive representative $A_{\mathrm{H}\alpha}$ in each redshift bin. Table~\ref{tab:CSFRD_dust} lists these $A_{\mathrm{H}\alpha}$ value alongside the corresponding CSFRD derived using that dust correction. We further discuss the \ha\ attenuation estimates in Section~\ref{discussion_dust}.

\section{\ha\ Luminosity Function results}\label{sec: LF}
\begin{figure*}[!t]
    \centering
    \includegraphics[width=\textwidth]{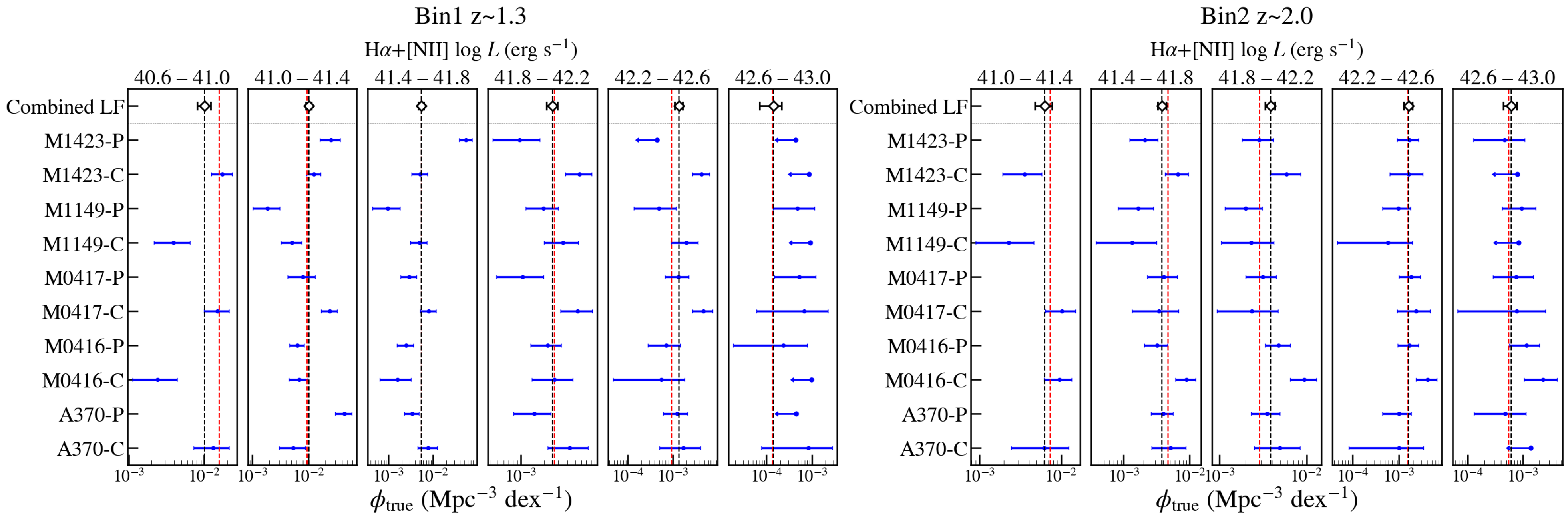}
    \caption{Observed \ha\ LFs of five lensing cluster fields and their parallel fields at $z\sim1.3$ (left panels) and $z\sim2.0$ (right panels). In each luminosity bin panel, blue points with error bars show the measured $\phi_\mathrm{true}$ values for individual fields (labeled on the left), with the suffixes ``-C" and ``-P" denoting cluster and parallel pointings, respectively. Fields lacking detections in the brightest bins are represented by upper limits. The white diamond at the top of each panel indicates the posterior median number densities obtained by combining the constraints from all fields in each luminosity bin. The black dashed vertical line marks this value, while the red dashed line indicates the LF value predicted by the best-fit Schechter function evaluated at the bin center.}
    \label{fig:luminosity_scatter}
\end{figure*}

We employed the $1/V_{\max}$ method \citep{1968ApJ...151..393S} to derive the \ha\ luminosity function (LF):
\begin{equation}
    \Phi(L) = \frac{1}{d\log L}\sum_i \frac{1}{C_i V_{\max,i}},
\end{equation}
where $C_i$ denotes the completeness correction for the $i$th source in a given luminosity bin, and $V_{\max,i}$ is the maximum comoving volume over which that source could be detected. The resulting LFs are shown as blue points in Figure~\ref{fig:luminosity_scatter}. The error bars incorporate both the Poisson uncertainty appropriate for small-number statistics \citep{1986ApJ...303..336G} and the uncertainty arising from cosmic variance, as described in Section~\ref{cosmic_variance}. For luminosity bins containing no detected \ha\ emitters, we report an upper limit rather than a zero-valued LF measurement. For these bins, we adopt a $1\sigma$ Poisson upper limit of 1.84 sources \citep{1986ApJ...303..336G}. We convert this count limit to an LF upper limit using $\Phi_{{\rm up},j} = \frac{1.84}{V_{{\rm eff},j}\,d\log L}$, 
where $V_{{\rm eff},j}$ is the effective accessible volume adopted for the empty bin. No cosmic-variance uncertainty is added to these zero-count limits. As all empty bins are at the bright end of the LF, we assume unit completeness when calculating their upper limits.

We fitted the \ha\ LFs derived from the five fields, including their corresponding parallel fields, with a Schechter function:
\begin{equation}
    \Phi(L) = \ln(10)\,\phi^{\ast}
    \left(\frac{L}{L^{\ast}}\right)^{\alpha+1}
    e^{-L/L^{\ast}},
\end{equation}
where $\phi^{\ast}$ represents the normalization of the luminosity function, $L^{\ast}$ is the characteristic luminosity, and $\alpha$ is the faint-end slope. We performed the fitting using a Markov Chain Monte Carlo (MCMC) approach. The prior ranges for the free parameters were set to $-3 < \alpha < -1$, and $40 < \log(L^{\ast}/{\rm erg\,s^{-1}}) < 45$. We adopted Gaussian priors for $L^{\ast}$ at $z\sim1.3$ and $z\sim2.0$, based on the measurements reported by \citet{2023ApJ...943....5N} and \citet{2013MNRAS.428.1128S}, respectively. \citet{2023ApJ...943....5N} fitted the LF without correcting for dust attenuation or [N\,II] contamination and applied attenuation of $A_{\mathrm{H}\alpha}=0.88$ mag when deriving SFRs, following a procedure similar to our analysis. The luminosities reported by \citet{2013MNRAS.428.1128S} were corrected for both effects, assuming $A_{\mathrm{H}\alpha}=1$ mag, equal to our fiducial value. The results from the joint fits combining all fields within each redshift bin are presented in Table~\ref{tab: LF_para}.

We also fitted the LF in each field independently. The resulting best-fit parameters are listed in Table~\ref{tab: LF_para_individual}. We find that the characteristic H$\alpha$ luminosity, $\log L_{\mathrm{H}\alpha}^{\ast}$, increases with redshift, while the faint-end slope, $\alpha_{\mathrm{H}\alpha}$, exhibits only weak redshift evolution. These trends are consistent with measurements at higher redshifts \citep{2025ApJ...987..186F}.

\begin{deluxetable}{lccccc}
\tablecaption{Galaxy counts and best-fit Schechter function parameters
for individual fields.
\label{tab: LF_para_individual}}

\tablehead{
\colhead{Field} &
\colhead{$N_{\rm gal}$} &
\colhead{H$\alpha$/H$\beta$} &
\colhead{$\log(L^\ast/{\rm erg\,s^{-1}})$} &
\colhead{$\log(\Phi^\ast/{\rm Mpc^{-3}dex^{-1}})$} &
\colhead{$\alpha$}
}
\startdata
\cutinhead{$1.1<z<1.5$}
ABELL-370 lens      & 28 & $5.69^{+0.47}_{-0.41}$ & $42.39^{+0.07}_{-0.07}$ & $-2.93^{+0.12}_{-0.08}$ & $-1.51^{+0.10}_{-0.08}$ \\
ABELL-370 parallel  & 47 & $3.90^{+0.36}_{-0.27}$ & $42.36^{+0.06}_{-0.07}$ & $-2.92^{+0.11}_{-0.08}$ & $-1.67^{+0.10}_{-0.08}$ \\
MACS-J0416 lens     & 28 & $4.39^{+0.31}_{-0.30}$ & $42.36^{+0.07}_{-0.08}$ & $-3.05^{+0.13}_{-0.08}$ & $-1.45^{+0.10}_{-0.08}$ \\
MACS-J0416 parallel & 52 & $3.35^{+0.19}_{-0.16}$ & $42.36^{+0.07}_{-0.08}$ & $-3.05^{+0.11}_{-0.08}$ & $-1.50^{+0.10}_{-0.08}$ \\
MACS-J0417 lens     & 61 & $3.24^{+0.13}_{-0.13}$ & $42.45^{+0.06}_{-0.07}$ & $-2.79^{+0.11}_{-0.07}$ & $-1.53^{+0.08}_{-0.07}$ \\
MACS-J0417 parallel & 30 & $4.73^{+0.25}_{-0.26}$ & $42.38^{+0.06}_{-0.07}$ & $-3.02^{+0.12}_{-0.08}$ & $-1.51^{+0.10}_{-0.08}$ \\
MACS-J1149 lens     & 42 & $3.77^{+0.17}_{-0.15}$ & $42.39^{+0.07}_{-0.08}$ & $-2.99^{+0.12}_{-0.08}$ & $-1.42^{+0.09}_{-0.07}$ \\
MACS-J1149 parallel & 25 & $3.73^{+0.22}_{-0.19}$ & $42.39^{+0.06}_{-0.07}$ & $-3.11^{+0.13}_{-0.08}$ & $-1.41^{+0.10}_{-0.09}$ \\
MACS-J1423 lens     & 82 & $4.07^{+0.17}_{-0.15}$ & $42.42^{+0.06}_{-0.07}$ & $-2.85^{+0.11}_{-0.07}$ & $-1.50^{+0.08}_{-0.06}$ \\
MACS-J1423 parallel & 38 & $4.05^{+0.34}_{-0.30}$ & $42.34^{+0.06}_{-0.07}$ & $-2.79^{+0.12}_{-0.08}$ & $-1.78^{+0.10}_{-0.08}$ \\
\cutinhead{$1.8<z<2.2$}
ABELL-370 lens      & 17 & $2.06^{+0.11}_{-0.09}$ & $42.72^{+0.11}_{-0.14}$ & $-3.19^{+0.14}_{-0.09}$ & $-1.54^{+0.15}_{-0.12}$ \\
ABELL-370 parallel  & 38 & $3.76^{+0.32}_{-0.28}$ & $42.70^{+0.11}_{-0.13}$ & $-3.21^{+0.13}_{-0.08}$ & $-1.53^{+0.15}_{-0.13}$ \\
MACS-J0416 lens     & 60 & $3.44^{+0.17}_{-0.16}$ & $42.87^{+0.10}_{-0.10}$ & $-3.04^{+0.13}_{-0.08}$ & $-1.50^{+0.11}_{-0.07}$ \\
MACS-J0416 parallel & 49 & $5.26^{+0.69}_{-0.54}$ & $42.79^{+0.09}_{-0.12}$ & $-3.15^{+0.13}_{-0.08}$ & $-1.47^{+0.14}_{-0.11}$ \\
MACS-J0417 lens     & 29 & $3.12^{+0.14}_{-0.13}$ & $42.77^{+0.10}_{-0.13}$ & $-3.17^{+0.14}_{-0.09}$ & $-1.50^{+0.13}_{-0.10}$ \\
MACS-J0417 parallel & 30 & $4.49^{+0.38}_{-0.32}$ & $42.77^{+0.10}_{-0.13}$ & $-3.17^{+0.13}_{-0.08}$ & $-1.49^{+0.15}_{-0.12}$ \\
MACS-J1149 lens     & 12 & $3.78^{+0.25}_{-0.23}$ & $42.67^{+0.12}_{-0.14}$ & $-3.27^{+0.15}_{-0.09}$ & $-1.37^{+0.15}_{-0.12}$ \\
MACS-J1149 parallel & 25 & $4.00^{+0.32}_{-0.26}$ & $42.76^{+0.10}_{-0.13}$ & $-3.24^{+0.14}_{-0.08}$ & $-1.33^{+0.15}_{-0.12}$ \\
MACS-J1423 lens     & 33 & $3.62^{+0.19}_{-0.17}$ & $42.73^{+0.10}_{-0.13}$ & $-3.16^{+0.13}_{-0.08}$ & $-1.48^{+0.12}_{-0.10}$ \\
MACS-J1423 parallel & 36 & $3.83^{+0.28}_{-0.23}$ & $42.76^{+0.10}_{-0.13}$ & $-3.21^{+0.13}_{-0.08}$ & $-1.38^{+0.14}_{-0.11}$ \\
\enddata
\tablecomments{\(N_{\rm gal}\) gives the number of galaxies
in the selected sample in each field and redshift bin. The H\(\alpha\)/H\(\beta\) column gives the values measured from
the stacked \ha\ and \hb\ emission-line maps, as described in Section~\ref{dust}.}
\end{deluxetable}

To summarize the LF constraints across regions, we constructed a joint posterior distribution for the number density in each luminosity bin. For regions with detections, we converted the measured LF and its uncertainties to $\log_{10}\Phi$ and combined them using piecewise Gaussian likelihoods in log space. Regions with zero detections contributed a Poisson likelihood term, $\exp(-\Phi V_{{\rm eff},i}\,d\log L)$. For bins containing at least one detection, we adopted a prior uniform in $\log_{10}\Phi$ and used the posterior median as the representative LF estimate, with uncertainties given by the equal-tailed 68.27\% credible interval, transformed back to linear LF units. 
The resulting representative LF estimates are listed in Table~\ref{tab: LF_para}.

\begin{figure*}[!t]
    \centering
    \includegraphics[width=\textwidth]{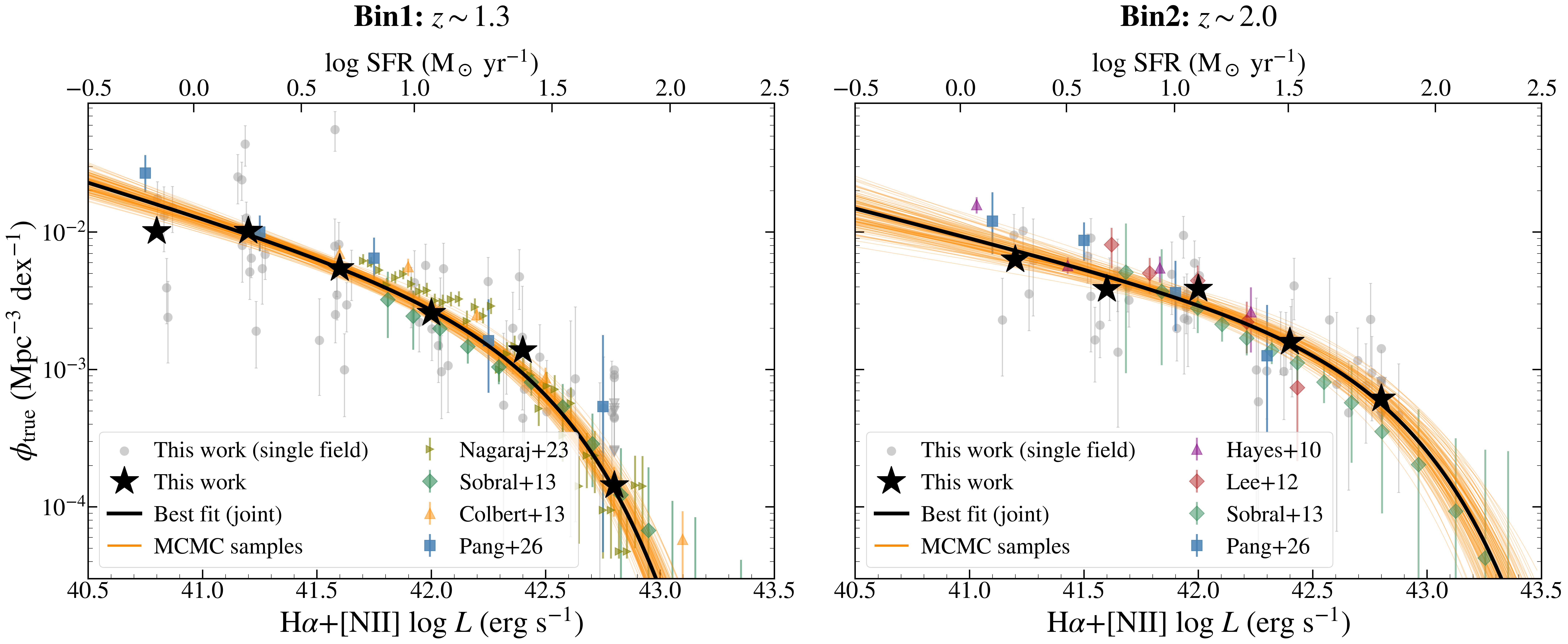}
    \caption{\ha\ LFs at $z\sim1.3$ (left) and $z\sim2.0$ (right).
    Light gray points show measurements in individual lensing cluster and parallel fields, while black stars indicate representative number densities obtained by combining all fields in each luminosity bin.
    Black curves show the best-fit Schechter functions, and light orange curves represent 100 random draws from the MCMC posterior.
    The upper axes show the corresponding SFRs, calculated using the [N \II]/\ha\ ratio from \citet{2018ApJ...855..132F} and the \ha-to-SFR conversion from \citet{2012ARA&A..50..531K}.
    Literature measurements include NIRISS spectroscopy of Abell~2744 (blue squares; P26), HST grism observations \citep[olive triangles;][]{2023ApJ...943....5N}, narrowband surveys \citep[purple triangles, green diamonds, and red diamonds, respectively;][]{2010A&A...509L...5H,2013MNRAS.428.1128S,2012PASP..124..782L}, and the WISP grism survey \citep[orange triangles;][]{2013ApJ...779...34C}.
    All literature measurements have been rescaled to a \citet{2003PASP..115..763C} IMF.
    Our measurements generally yield lower faint-end number densities than previous studies, particularly at $z\sim1.3$.
    }
    \label{fig:LF_results}
\end{figure*}

These binwise posterior summaries are displayed as open diamonds connected by black dashed lines at the top of each panel in Figure~\ref{fig:luminosity_scatter} and as black stars in Figure~\ref{fig:LF_results}. 
In Figure~\ref{fig:LF_results}, the black curves show the best-fit Schechter functions, while the orange curves illustrate a subset of the MCMC posterior samples. The best-fit functions evaluated at the luminosity-bin centers are also marked by red vertical lines in Figure~\ref{fig:luminosity_scatter}, allowing comparison with both the individual-field measurements and the representative LF estimates.

We note that the bright end of the LF is not well constrained by our observations due to the limited survey volume and the small number of luminous sources. Therefore, the bright-end behavior of our fitted LFs, particularly at $z\sim1.3$, is primarily constrained by the adopted priors.

We adopted the calibration from \citet{2012ARA&A..50..531K} to derive the SFR from the H$\alpha$ + [N\,II] emission:
\begin{equation}
    \log\left(\frac{\mathrm{SFR}}{M_\odot\,\mathrm{yr}^{-1}}\right)
    =
    \log\left(\frac{L_{\mathrm{H}\alpha}}{\mathrm{erg\,s}^{-1}}\right)
    -41.27+\log(0.94).
\end{equation}
We assumed that our galaxies follow the star-forming main-sequence (MS) relation derived by \citet{speagle2014highly}:
\begin{equation}
\begin{aligned}
    \log\left(\frac{\mathrm{SFR}}{M_\odot\,\mathrm{yr}^{-1}}\right)
    =& (0.84-0.026t)\log\left(\frac{M_*}{M_\odot}\right) \\
    &-(6.51-0.11t).
\end{aligned}
\end{equation}
where $t$ represents the age of the Universe in Gyr. By combining this relation with the stellar mass--[N\,II]/H$\alpha$ ratio relation from \citet{2018ApJ...855..132F},
\begin{equation}
\begin{aligned}
    \log\left(\frac{M_*}{M_\odot}\right)
    =& 3.696\xi +3.236\xi^{-1}+0.729\xi^{-2} \\
    &+14.928+0.156(1+z)^2 .
\end{aligned}
\end{equation}
where
\begin{equation}
    \xi = \log\left(\frac{[\mathrm{N\,II}]}{\mathrm{H}\alpha}\right)
    +0.138-0.042(1+z)^2 ,
\end{equation}
and applying the dust corrections described in Section~\ref{dust}, we corrected the observed H$\alpha$ + [N\,II] fluxes for [N\,II] contamination and dust attenuation and converted them into star formation rates. The upper x-axis in Figure~\ref{fig:LF_results} shows the corresponding SFR assuming a fixed attenuation of $A_{\mathrm{H}\alpha}=1\,\mathrm{mag}$.


\begin{deluxetable}{cccc}
\tablewidth{300\%}
\tablecaption{\centering Schechter Fit Parameters of the H$\alpha$ Luminosity Functions
\label{tab: LF_para}}
\tabletypesize{\footnotesize}
\tablehead{
\colhead{$z$} & \colhead{$\log L^{*}_{\rm H\alpha}$} & \colhead{$\log \Phi^{*}_{\rm H\alpha}$} & \colhead{$\alpha_{\rm H\alpha}$} \\
  & ($\rm erg~s^{-1}$) & \colhead{($\rm Mpc^{-3}~dex^{-1}$)} & 
}
\startdata
$1.1<z<1.5$ & $42.41_{-0.05}^{+0.04}$ & $-2.97_{-0.06}^{+0.08}$ & $-1.51_{-0.04}^{+0.06} $ \\
$1.8<z<2.2$ & $42.80_{-0.09}^{+0.06}$ & $-3.16_{-0.07}^{+0.13}$ & $-1.42_{-0.06}^{+0.09} $ \\
\enddata
\end{deluxetable}

\section{Discussion} \label{sec: discussion}
\subsection{Cosmic Variance} \label{cosmic_variance}
We utilized the Cosmic Variance Calculator developed by \citet{2008ApJ...676..767T} to estimate the cosmic variance uncertainty for each field in each luminosity bin and redshift interval.

To determine the appropriate filling factor, we referred to the results presented by \citet{2011ApJ...731..113M}. For each redshift bin, we selected a representative luminosity bin in one field and estimated the expected number of detected sources within the same survey area as \citet{2011ApJ...731..113M}, assuming that the number of sources scales proportionally with the effective survey volume at the same redshift and sky position. We then varied the filling factor in the Cosmic Variance Calculator developed by \citet{2008ApJ...676..767T} until the resulting cosmic variance was consistent with the estimate from \citet{2011ApJ...731..113M}. The derived filling factors at different redshifts were subsequently adopted to calculate the cosmic variance uncertainties for all fields and luminosity bins. The relative cosmic variance is approximately $20\%-30\%$ for individual fields.

Since our survey comprises ten independent fields, the contribution from cosmic variance is reduced when combining measurements across all fields, following Equation~(7) of \citet{2011ApJ...731..113M}. The relative cosmic variance of the combined sample decreases to $\sim0.1$, about 2 to 4 times smaller than that of the individual fields. 
This demonstrates the advantage of combining multiple independent sightlines in mitigating the impact of cosmic variance on the LF measurements.

During the LF fitting, we incorporated the cosmic variance uncertainties into the Poisson uncertainties of the number densities measured in each field. Figure~\ref{fig:luminosity_scatter} shows the LFs derived for all individual fields included in the fitting, together with the mean number density in each luminosity bin. The measurements exhibit substantial field-to-field scatter, highlighting the impact of cosmic variance. By combining multiple independent fields and probing a larger effective survey volume, our LF measurements are less sensitive to field-to-field fluctuations, thereby reducing the uncertainty associated with cosmic variance.

\subsection{Faint-end Slope of LF} \label{subsec: Faint-end}
\begin{figure} 
    \includegraphics[width=0.5\textwidth]{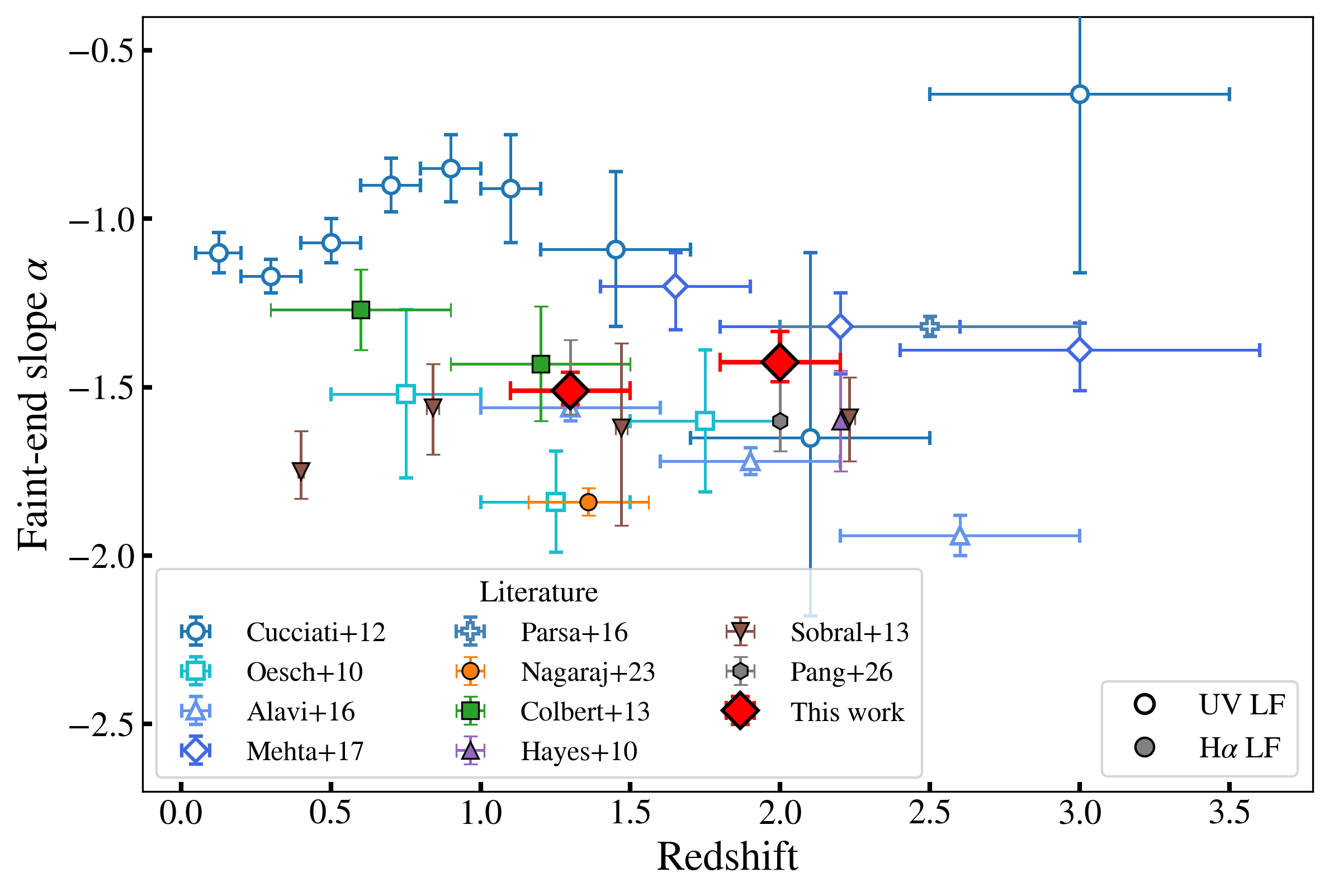}
    \caption{Evolution of the faint-end slope, $\alpha$, of the H$\alpha$ and rest-frame UV luminosity functions with redshift. The filled symbols show measurements from H$\alpha$ luminosity functions, while the open symbols represent measurements from UV luminosity functions. Different markers indicate results from different studies. The red diamonds show the measurements from this work at $z\sim1.3$ and $z\sim2.0$. Literature measurements of the H$\alpha$ faint-end slope are taken from \citet{2023ApJ...943....5N}, \citet{2013ApJ...779...34C}, \citet{2010A&A...509L...5H}, \citet{2013MNRAS.428.1128S}, and P26, while the UV measurements are taken from \citet{2012A&A...539A..31C}, \citet{2010ApJ...725L.150O}, \citet{2016ApJ...832...56A}, \citet{2017ApJ...838...29M}, and \citet{2016MNRAS.456.3194P}.}
    \label{fig:alpha}
\end{figure}

In this work, we extend the H$\alpha$ LF to luminosities as faint as $\sim0.01L^\ast$ using grism spectroscopy, enabling robust constraints on the faint-end slope, $\alpha_{\mathrm{H}\alpha}$. We measure $\alpha_{\mathrm{H}\alpha}=-1.51^{+0.06}_{-0.04}$ at $z\sim1.3$ and $\alpha_{\mathrm{H}\alpha}=-1.42^{+0.09}_{-0.06}$ at $z\sim2.0$. In Figure~\ref{fig:alpha}, we compare our measurements with previous determinations of the faint-end slopes of the H$\alpha$ and rest-frame UV LFs. At $z\sim1.3$, our measurement is broadly consistent with previous H$\alpha$ studies, whereas at $z\sim2.0$, we find a shallower faint-end slope than most previous determinations. Our best-fitting values suggest a mild flattening of the H$\alpha$ LF with increasing redshift over this interval. However, the two measurements are consistent within their uncertainties.

We note that some UV LF studies report significantly flatter faint-end slopes, particularly \citet{2012A&A...539A..31C}. One possible contributing factor is the relatively limited depth of their data, which provides less leverage on the faint galaxy population and may enhance degeneracies among the Schechter-function parameters. The faint-end absolute-magnitude limits adopted by \citet{2012A&A...539A..31C} are approximately $M_{\rm UV}=-18.4$ at $1.2<z<1.7$ and $M_{\rm UV}=-19.8$ at $1.7<z<2.5$, corresponding to unobscured SFRs of $\sim1.4$ and $\sim5.1\,M_\odot\,{\rm yr}^{-1}$, respectively, substantially higher than the SFR limits probed in this work. However, \citet{2012A&A...539A..31C} performed several tests for selection effects and incompleteness and still recovered similarly flat slopes. Therefore, while the limited faint-end coverage may weaken the constraints on $\alpha$, we cannot exclude the possibility that the flat slope reflects an intrinsic feature of their measured UV LF rather than being solely driven by sample incompleteness.

Apart from observational depth, the relative faint-end slopes of the H$\alpha$ and UV LFs can also be affected by dust attenuation and variations in the H$\alpha$-to-UV luminosity ratio. Rest-frame UV emission is more strongly attenuated by dust than H$\alpha$ emission \citep{1999PASP..111...63F, calzetti2000}, while the amount of dust attenuation is known to correlate with stellar mass and SFR \citep{2020A&A...644A.144D}. As discussed by \citet{2012ApJ...758...93F}, these effects can result in a UV LF that is steeper than the dust-corrected H$\alpha$ LF.

On the other hand, variations in the $L_{\mathrm{H}\alpha}/L_{\mathrm{UV}}$ ratio can act in the opposite direction. The $L_{\mathrm{H}\alpha}/L_{\mathrm{UV}}$ ratio depends on the recent star-formation activity and physical properties of galaxies \citep{2025ApJ...987..186F}. Low-metallicity galaxies tend to have higher ionizing-photon production efficiencies \citep{2019MNRAS.486.2215K}, while low-mass galaxies are expected to experience more bursty star-formation histories \citep{2014ApJ...792...99S,2019ApJ...881...71E}. Consequently, galaxies with lower UV luminosities can exhibit systematically higher $L_{\mathrm{H}\alpha}/L_{\mathrm{UV}}$ ratios \citep{2017MNRAS.465.3637M, 2020MNRAS.493.5120M, 2022MNRAS.511.4464A}, potentially producing a steeper faint-end slope for the H$\alpha$ LF than for the UV LF. Differences in the assumed IMF may introduce additional systematic variations in the inferred LF shapes \citep{2008ApJ...675..163H}.

We note that our measurements exhibit substantial field-to-field variation (see Figure~\ref{fig:luminosity_scatter}). To assess its impact on the LF fitting, we also fit the luminosity function for each field individually, with the resulting best-fit parameters listed in Table~\ref{tab: LF_para_individual}. The inferred faint-end slopes show considerable scatter, ranging from $-1.78$ to $-1.33$. We also caution that parameter degeneracies in the Schechter-function fitting may affect the faint-end slope inferred from the combined sample.

\subsection{Dust Attenuation in the Sample} \label{discussion_dust}
We use the stacked Balmer decrements described in Section~\ref{dust} to constrain dust attenuation in our sample. According to table \ref{tab: LF_para_individual}, the ABELL-370 lensing field at $z\sim2.0$ shows H$\alpha$/H$\beta$ ratios below the intrinsic Case B recombination value of 2.86. 
This low ratio suggests little or no dust attenuation, with observational uncertainties potentially scattering the measured values below 2.86. Emission-line selection may preferentially select actively star-forming galaxies with low dust attenuation \citep{acharyya2026spatially}. Similar Balmer decrements below the standard Case B value have also been reported in JWST/NIRISS observations \citep{pirzkal2024next}. Balmer-line radiative-transfer effects or density-bounded nebulae offer additional possible explanations \citep{scarlata2024universal,mcclymont2025density}. 
We therefore caution that the derived $A_{\mathrm{H}\alpha}$ values should be interpreted statistically rather than as precise measurements of the dust attenuation in each field. 

Moreover, at $z\sim2$, we generally infer lower $A_{H\alpha}$ values for the lensing fields than for the corresponding parallel fields (see Table~\ref{tab:CSFRD_dust}). One possible contributor to this difference is the intrinsic luminosities in the two sets of fields. Gravitational magnification allows intrinsically fainter galaxies to enter the lensing-field samples; such galaxies may typically have lower dust attenuation \citep{sobral2012star}. 

We compared our results with the previous spatially resolved measurements of \citet{2023ApJ...949L..11M}. Using CANUCS JWST/NIRISS spectroscopy, they studied dust attenuation by stacking galaxies from the MACS-J0417 cluster and parallel field in stellar mass and redshift bins. Within the central 0.4~kpc, they measured $A_{\mathrm{H}\alpha}=0.93\pm0.05$ and $1.03\pm0.04\,\mathrm{mag}$ for galaxies with $7.6\leq\log(M_\star/M_\odot)<9.0$ at $1.0\leq z<1.7$ and $1.7\leq z<2.4$, respectively. For $9.0\leq\log(M_\star/M_\odot)<10.0$, the corresponding values are $0.79\pm0.03$ and $1.07\pm0.04\,\mathrm{mag}$.

\citet{2023ApJ...949L..11M} stacked galaxies from both MACS-J0417 pointings together. Their central $A_{\mathrm{H}\alpha}$ estimates fall between the values we infer separately for the lensing and parallel fields, suggesting broad agreement (see Table~\ref{tab: LF_para_individual}). Their estimates refer to the central $0.4\,\mathrm{kpc}$, whereas ours include emission from the outer regions of the galaxies. \citet{2026ApJ..1005L..20R} found that the $A_{\mathrm{H}\alpha}$ declines with radius in galaxies lensed by Abell~2744 at both $z\sim1.3$ and $z\sim2$. If a similar trend holds in our sample, emission from less attenuated outer regions could lower the integrated $A_{\mathrm{H}\alpha}$ relative to a measurement within the central $0.4\,\mathrm{kpc}$.

\subsection{Estimate of Cosmic SFR Density} \label{subsec: CSFRD}
We estimated the cosmic star formation rate density (CSFRD, $\rho_{\mathrm{SFR}}$) at $z\sim1.3$ and $z\sim2.0$ by integrating the star formation rate functions (SFRFs) derived from the best-fit Schechter functions. The lower integration limit was set to $0.4\,M_\odot\,\mathrm{yr}^{-1}$, corresponding to an H$\alpha$ luminosity of $L_{\mathrm{H}\alpha}=10^{40.5}\,\mathrm{erg\,s}^{-1}$ and an attenuated absolute UV magnitude limit of $M_{\mathrm{UV}}=-17.6$ mag. To estimate the uncertainties, we performed Monte Carlo sampling of the posterior distributions of the SFRF parameters. For each posterior realization, we integrated the corresponding SFRF to obtain the 16th, 50th, and 84th percentiles of $\rho_{\mathrm{SFR}}$.

\begin{figure*}[!t]
\centering
    \includegraphics[width=\textwidth]{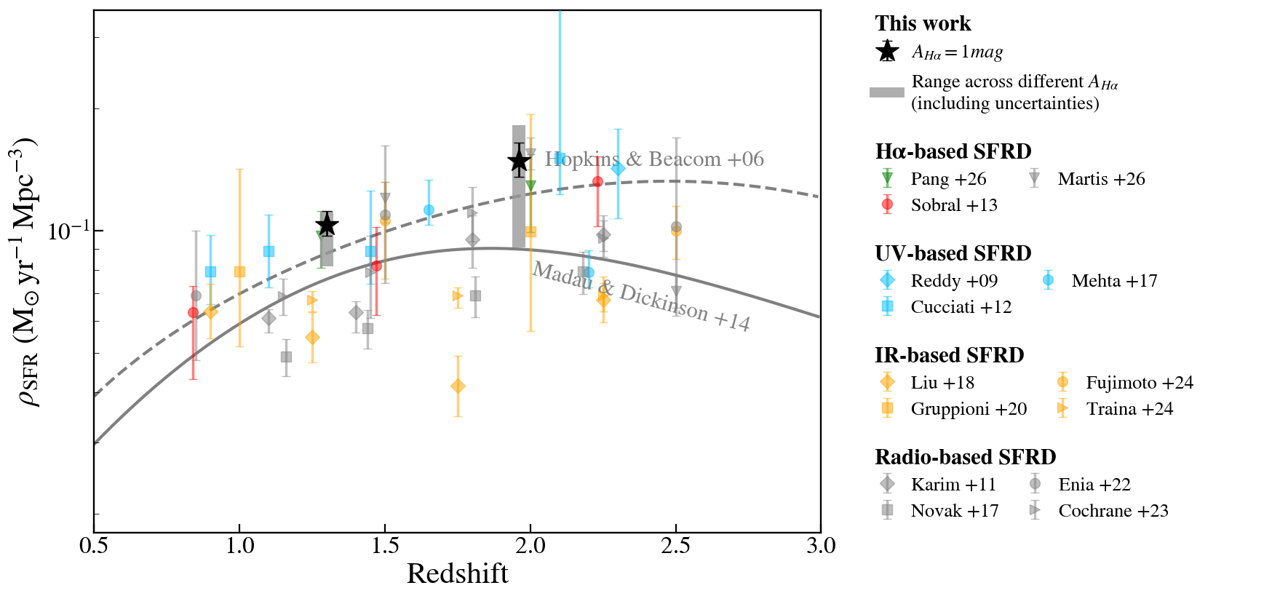}
    \vspace*{-1em}
    \caption{The cosmic star formation rate density (SFRD) from $z \sim 0.5$ to $z \sim 3.0$. Our measurements assuming a common attenuation of $A_{\mathrm{H}\alpha}=1\ \mathrm{mag}$ in both redshift bins are shown as black stars, with error bars indicating the corresponding SFRD uncertainties. 
    The gray vertical ranges span the CSFRD estimates and their uncertainties for all adopted $A_{\mathrm{H}\alpha}$ values. For comparison, we include literature estimates from H$\alpha$ surveys \citep{Pang_, 2013MNRAS.428.1128S,martis2026new}, dust-corrected UV LFs \citep{2009ApJ...692..778R, 2012A&A...539A..31C, 2017ApJ...838...29M}, infrared-based studies \citep{2018ApJ...853..172L, 2020A&A...643A...8G, 2024ApJS..275...36F, 2024A&A...681A.118T}, and radio-based measurements \citep{2011ApJ...730...61K, 2017A&A...602A...5N, 2022ApJ...927..204E, 2023MNRAS.523.6082C}. The solid and dashed curves represent the analytical fits from \citet{2014ARA&A..52..415M} and \citet{2006ApJ...651..142H}, respectively. All literature values have been rescaled to a \citet{2003PASP..115..763C} IMF for consistency.
    }
    \label{fig:CSFRD}
\end{figure*}

Assuming a common attenuation of $A_{\mathrm{H}\alpha}=1\,\mathrm{mag}$ in both redshift bins, we derive $\rho_{\rm SFR}=0.103^{+0.008}_{-0.007}\,M_\odot\,{\rm yr}^{-1}\,{\rm Mpc}^{-3}$ at $z\sim1.3$ and $\rho_{\rm SFR}=0.149^{+0.016}_{-0.013}\,M_\odot\,{\rm yr}^{-1}\,{\rm Mpc}^{-3}$ at $z\sim2.0$, tracing the evolution of the CSFRD at fixed attenuation. 
We examined the sensitivity of our CSFRD estimates to the adopted dust attenuation correction. 
We recalculated the CSFRDs by adopting the attenuation values from \citet{2023ApJ...949L..11M} and our stacked results as alternative corrections within each redshift bin. The resulting CSFRDs are summarized in Table~\ref{tab:CSFRD_dust} and shown in Figure~\ref{fig:CSFRD}. 

Our SFRD measurements are broadly consistent with previous H$\alpha$-based and dust-corrected UV estimates \citep{Pang_,2013MNRAS.428.1128S,martis2026new,2009ApJ...692..778R,2012A&A...539A..31C,2017ApJ...838...29M}, although our estimate at $z\sim2.0$ is slightly higher than that of P26. Our measurements also lie above the empirical fits of \citet{2006ApJ...651..142H} and \citet{2014ARA&A..52..415M}, shown as the dashed and solid gray curves in Figure~\ref{fig:CSFRD}, respectively. For consistency, literature SFRD measurements have been converted to a \citet{2003PASP..115..763C} IMF using scaling factors of 0.63 and 0.94 for the \citet{1955ApJ...121..161S} and \citet{kroupa2001variation} IMFs, respectively. 

The H$\alpha$- and UV-based measurements compared in Figure~\ref{fig:CSFRD} adopt different dust attenuation prescriptions. Both \citet{2013MNRAS.428.1128S} and P26 apply a uniform correction of $A_{\mathrm{H}\alpha}=1\,\mathrm{mag}$, consistent with our fixed-attenuation calculation. In contrast, \citet{martis2026new} correct individual galaxies using stellar attenuation estimates from SED fitting. Among the UV studies, \citet{2009ApJ...692..778R} adopt a luminosity-dependent reddening distribution with the \citet{calzetti2000} curve, whereas \citet{2012A&A...539A..31C} apply a redshift-dependent mean FUV attenuation inferred from SED fitting. \citet{2017ApJ...838...29M} combine a luminosity-dependent UV-slope distribution with IRX--$\beta$ relations. The adopted \citet{2009ApJ...692..778R} estimate additionally includes an infrared-based contribution from highly luminous, dust-obscured galaxies. These differences in attenuation estimates, their dependence on galaxy properties, and the treatment of obscured populations can contribute to the spread in the inferred CSFRDs even after conversion to a common IMF. The differences among these estimates illustrate the sensitivity of the inferred CSFRD to the adopted dust attenuation correction. CSFRD estimates derived from less dust-sensitive tracers, such as Pa$\alpha$ LF, could help reduce the uncertainties and provide more robust constraints on the cosmic star formation history.

\begin{deluxetable}{ccl}[t]
\tabletypesize{\footnotesize}
\tablewidth{\columnwidth}
\tablecolumns{3}
\setlength{\tabcolsep}{3pt}
\tablecaption{CSFRDs derived using different dust attenuation values.\label{tab:CSFRD_dust}}
\tablehead{
\colhead{$A_{\mathrm{H}\alpha}$} &
\colhead{$\rho_{\rm SFR}$} &
\colhead{Attenuation source} \\
\colhead{$(\mathrm{mag})$} &
\colhead{$(M_\odot\,\mathrm{yr}^{-1}\,\mathrm{Mpc}^{-3})$} &
\colhead{}
}
\startdata
\cutinhead{$z \sim 1.3$}
$1.00$         & $0.103^{+0.008}_{-0.007}$ & \\
$0.89\pm0.06$  & $0.095^{+0.008}_{-0.006}$ & This work: central cluster fields \\
$0.82\pm0.09$  & $0.089^{+0.007}_{-0.006}$ & This work: parallel fields \\
$0.93\pm0.05$  & $0.098^{+0.008}_{-0.006}$ & M23: low mass \\
$0.79\pm0.03$  & $0.087^{+0.007}_{-0.006}$ & M23: intermediate mass \\
\cutinhead{$z \sim 2.0$}
$1.00$         & $0.149^{+0.016}_{-0.013}$ & \\
$0.51\pm0.08$  & $0.100^{+0.011}_{-0.009}$ & This work: central cluster fields \\
$1.12\pm0.15$  & $0.164^{+0.018}_{-0.014}$ & This work: parallel fields \\
$1.03\pm0.04$  & $0.152^{+0.016}_{-0.013}$ & M23: low mass \\
$1.07\pm0.04$  & $0.157^{+0.017}_{-0.014}$ & M23: intermediate mass \\
\enddata
\tablecomments{
M23 refers to \citet{2023ApJ...949L..11M}.
The low- and intermediate-mass bins correspond to
$7.6\leq\log(M_\star/M_\odot)<9.0$ and
$9.0\leq\log(M_\star/M_\odot)<10.0$, respectively.
}
\end{deluxetable}

Differences in the adopted integration limits may contribute to these offsets. \citet{2014ARA&A..52..415M} integrated the luminosity functions down to $0.03L^\ast$, whereas we adopt a fixed lower limit of $\log_{10}[L_{\mathrm{H}\alpha}/(\mathrm{erg\,s^{-1}})]=40.5$ in both redshift bins. Given our measured characteristic luminosities (Table~\ref{tab: LF_para}), our limits are $0.39$ and $0.78$~dex fainter than $0.03L^\ast$. For the compilation of \citet{2006ApJ...651..142H}, the luminosity functions were extrapolated to zero luminosity where possible, but a uniform integration limit was not applied to all measurements; some high-redshift estimates were integrated only down to $0.1$--$0.2L^\ast$ \citep{li2008star}. To quantify the sensitivity of our results to the integration limit, we repeated the calculation using the same relative threshold, $0.03L^\ast$, as \citet{2014ARA&A..52..415M}. This yields $\rho_{\mathrm{SFR}}=0.094$ and $0.131\,M_\odot\,\mathrm{yr}^{-1}\,\mathrm{Mpc}^{-3}$ at $z\sim1.3$ and $z\sim2.0$, respectively, bringing our measurements into closer agreement with the \citet{2006ApJ...651..142H} fit. This comparison illustrates the effect of the integration limit, although differences in SFR tracers and calibrations also need to be considered. 


Both H$\alpha$ and rest-frame UV emission ultimately arise from young, massive stars. While H$\alpha$ traces the ionizing photons produced by the most massive and short-lived stars, the UV continuum remains sensitive to young stellar populations over somewhat longer timescales. The bright ends of both the UV and H$\alpha$ LFs are generally well described by the exponential decline \citep{2013MNRAS.428.1128S,martis2026new,2009ApJ...692..778R,2012A&A...539A..31C,2017ApJ...838...29M}.

In contrast, radio and far-infrared (FIR) emission provide more indirect tracers of star formation. FIR emission primarily arises from dust heated by absorbed stellar radiation. Moreover, FIR emission does not necessarily originate exclusively from active star-forming regions, as dust can also be heated by older stellar populations, diffuse interstellar radiation fields, and AGN activity \citep{1987ApJ...314..513L,2009ApJ...700..161S}. The bright ends of radio and FIR LFs are often better characterized by a shallower, power-law-like decline rather than a simple exponential cutoff \citep{2018ApJ...853..172L,2020A&A...643A...8G,2024ApJS..275...36F,2024A&A...681A.118T,2011ApJ...730...61K,2017A&A...602A...5N,2022ApJ...927..204E,2023MNRAS.523.6082C}. In addition, previous radio- and FIR-based studies adopt a wide range of lower integration limits (see Table~4 of P26), introducing an additional source of systematic uncertainty when comparing their inferred SFRDs.

\section{Summary} \label{sec: summary}
In this work, we identified \ha\ emitters in five lensing fields and their parallel fields using JWST/NIRISS grism slitless spectroscopy from the CANUCS program. Our main results and conclusions are summarized as follows.

\begin{enumerate}
    \item We constructed a sample of 433 \ha\ emitters at $1.1<z<1.5$ and 329 \ha\ emitters at $1.8<z<2.2$. We identified one Type \I\ AGN with a broad \ha\ component in the MACS-J0417 field and nine multiply imaged systems across the five lensing fields and their parallel fields.
    
    \item Following the methodology of P26, we estimated the survey volume, completeness, and Eddington bias for each source, and quantified the \ha\ emitter number density in each field. We further incorporated cosmic variance into the uncertainty estimate. The significant field-to-field variation in the number density highlights the importance of sampling multiple independent sightlines in galaxy population studies.
    
   \item We measured the H$\alpha$ luminosity functions at $z\sim1.3$ and $z\sim2.0$, reaching luminosities as faint as $\log L_{\mathrm{H}\alpha}\sim40.6$ and $\sim41.0$, respectively. The inferred faint-end slopes are $\alpha_{\mathrm{H}\alpha}=-1.51^{+0.06}_{-0.04}$ and $-1.42^{+0.09}_{-0.06}$, respectively. The measured faint-end slopes suggest a trend toward shallower slopes with increasing redshift. At $z\sim1.3$, our slope is consistent with previous H$\alpha$ LF measurements, whereas at $z\sim2.0$, it is slightly shallower than those reported in previous H$\alpha$ LF studies. Individual fields exhibit a broad range of faint-end slopes. These results underscore the importance of field-to-field variations and degeneracies among the Schechter parameters when interpreting faint-end slope measurements.

    \item We derived the CSFRD in both redshift bins using multiple dust attenuation estimates, including those obtained with the stacking-based approach of \citet{2026ApJ..1005L..20R} and the $A_{\mathrm{H}\alpha}$ values reported by \citet{2023ApJ...949L..11M}. Our CSFRD estimates are consistent with previous H$\alpha$-based measurements at $z\sim1.3$ but are slightly higher at $z\sim2.0$. The inferred CSFRD depends sensitively on the adopted dust attenuation correction, highlighting dust corrections as an important source of systematic uncertainty in H$\alpha$-based CSFRD measurements.
    The higher CSFRD at $z\sim2$ than at $z\sim1.3$ is consistent with the overall rise of cosmic star formation activity toward its peak around Cosmic Noon.
\end{enumerate}
Our study expands the sample presented by P26 and provides tighter constraints on the \ha\ luminosity function during Cosmic Noon. We demonstrate the capability of JWST/NIRISS grism spectroscopy in measuring luminosity functions of emission-line galaxies, and further validate the methodology for estimating survey volume, completeness, and Eddington bias introduced by P26.

The two-dimensional grism data of our sample, which provide reliable redshift measurements and clear emission-line detections, also offer valuable opportunities for future studies of spatially resolved galaxy properties, such as dust attenuation \citep{2026ApJ..1005L..20R}. The JWST data used in this work are publicly available through the CANUCS/Technicolor Data Release 1:
\url{https://niriss.github.io/data_release1.html}.

\begin{acknowledgments}
We thank Chao-Wei Tsai for insightful and constructive comments on our dust attenuation analysis. This work is supported by Beijing Natural Science Foundation (QY26525), the National Natural Science Foundation of China (grant 12373009), the National Key R\&D Program of China No.2025YFF0510603, the CAS Project for Young Scientists in Basic Research Grant No. YSBR-062, the China Manned Space Program with grant no. CMS-CSST-2025-A06, and the Fundamental Research Funds for the Central Universities. XW acknowledges the support by the Xiaomi Young Talents Program, and the work carried out, in part, at the Swinburne University of Technology, sponsored by the ACAMAR visiting fellowship.
PJW acknowledges support from the Agence Nationale de la Recherche (ANR), grant ANR-25-CE31-0192-01 (CIGALE-2D). 
\end{acknowledgments}


\bibliography{sample701}{}
\bibliographystyle{aasjournalv7}



\end{document}